\documentclass[fleqn,14pt]{article}
\usepackage{amssymb}
\usepackage{amsmath}
\usepackage{color}
\usepackage{geometry}
\newcommand{\be}{\begin{equation}}
\newcommand{\ee}{\end{equation}}

\begin{document}

\begin{center}
{\Large FUNDAMENTAL PROPERTIES OF QUATERNION SPINORS}\\[1em]
\small Alexander P. Yefremov\\\textsl{a.yefremov@rudn.ru} \\[0.5em]
Peoples' Friendship University of Russia, Miklukho-Maklaya St. 6, Moscow, Russia\\[2em]
\textbf{Abstract}
\end{center}
{\small
Interior structure of arbitrary sets of quaternion units is analyzed using general methods of the theory of matrices. It is shown that the units are composed of quadratic combinations of fundamental objects having dual mathematical meaning as spinor couples and dyads locally describing 2D-surfaces. A detailed study of algebraic interrelations between the spinor sets belonging to different quaternion units is suggested as an initial step aimed to produce a self-consistent geometric image of spinor-surface distribution on the physical 3D space background.\\[1em]
%
%Анализируется внутренняя структура произвольных множеств кватернионных единиц с использованием общих методов теории матриц. Показано, что данные единицы составлены из квадратичных комбинаций фундаментальных объектов, имеющих двойной математический смысл: спинорные пары и диады, локально описывающие 2-мерные поверхности. Проведен подробный анализ алгебраических взаимосвязей между множествами спиноров, принадлежащих к различным кватернионным единицам, как первый шаг к получению самосогласованной геометрической картины распределения спинорных поверхностей на фоне физического 3-мерного пространства.}\\[1em]

\section{Introduction: Physical domains in quaternion mathematics}

Study of hypercomplex numbers reveals many mathematical correlations resembling formulas of physical theories known from experiment or heuristic assumptions. These coincidences are found in the associative in multiplication algebras of double (split-complex) numbers, dual numbers, quaternion numbers, and of embracing all them bi-quaternion algebra, so the basement of all these algebras are four quaternion (Q-) units [1]. Hamilton, the inventor of quaternions, was the first to discover that the triad of imaginary Q-units behaves as a set of unit vectors initiating the orthogonal system of coordinates in 3D space, a geometric and physical fact [2]; Descartes introduced this important object heuristically more than 200 years before Hamilton's quaternions. Maxwell also used quaternions to formulate the equations of electrodynamics, and 70 years later Fueter strikingly found that these equations (in vacuum) are equivalent to differentiability conditions for a vector function of Q-variable [3]. The spin-term introduced by Pauli into Schr\"odinger equation, later was shown to reflect Q-properties of the quantum-mechanical space [4]. Further development of the notion of Q-space exposed formal equivalence of the space curvature and the Yang-Mills field intensity [5], while Q-description of the movable frames resulted in specific (rotational) version of the theory of relativity [6,7]. An important property of Q-numbers is linked with spinors. Rastall [8] identified spinors with Q-ideals, mathematical objects emerging in equations involving idempotent matrices, but this observation was in fact used only to offer a different description of the Dirac spinor theory. 

It seems though that the Q-spinors deserve more detailed study, since they not only can serve as material for an alternative reconstitution of known physical structures, but themselves form a powerful instrument for revealing of the space (and time) properties. This study is aimed to widely analyze the origin of the quaternion spinors, the freedoms they dispose, their deep links with the structure of Q-numbers and their distribution on the 3D space background. In Section 2 a sketch of quaternion algebra is given with samples of the representations of the algebra units. In Section 3 a component-free procedure based on the theory of matrices is developed for detecting spinors in the structure of Q-units. In Section 4 the relations linking spinors associated with different directions (dimensions) of a Q-space are deduced. Section 5 contains complete tables of mutual projections of spinor-vectors, and discussion in Section 6 concludes the study.

\section{Quaternions in short}

A quaternion $a=x\cdot I+y_1 \cdot {\bf q}_1 +y_2 \cdot {\bf q}_2 +y_3 \cdot {\bf q}_3$ is a number built on the basis of one real (scalar) unit \textit{I} and three imaginary (vector) units\footnote{ This vector notations relate to traditional Hamilton's notations (still used in literature) as
${\bf q}_1 =\textbf{i},\; {\bf q}_2 =\textbf{j},\; {\bf q}_3 =\textbf{k}$.}
${\bf q}_{1}, {\bf q}_{2}, {\bf q}_{3}$, each unit having real coefficients $\{x, y_1, y_2, y_3 \} \in \mathbb{R}$. Let small Latin indices enumerate vector units $k, m, n,\, ...=1, 2, 3$, then the quaternion is rewritten in a compact form $a=x+y_{k} {\bf q}_{k} $, summation in the repeated indices is implied, the real unit symbol is traditionally omitted. Q-numbers admit algebraic addition similar to that of complex numbers, multiplication is associative but no more commutative, the vector units commute with the real unit but do not commute between themselves; this is reflected by the multiplication table 
\be\label{eq1}
I{\bf q}_{k} ={\bf q}_{k} I={\bf q}_{k},\;\;
{\bf q}_{k}{\bf q}_{n} =-\delta_{kn} +\varepsilon_{knm}{\bf q}_{m},
\ee
where $\delta_{kn}$, $\varepsilon_{knm}$ are 3D Kronecker and Levi-Civita symbols. Q-conjugation $\bar{a}=x-y_k {\bf q}_k$ helps to define the modulus of a Q-number 

\[
|a|=\sqrt{a\bar{a}}=\sqrt{x^2+y_k y_k} \equiv \sqrt{x^2+y^2} \in \mathbb{R}
\] 
with two consequences. First, inverse quaternion is defined 
\[
a^{-1} =\frac{\bar{a}}{|a|^{2} },
\] 
so that quaternion division (right and left as multiplication) can be introduced 
\[\left(\frac{a}{b} \right)_\text{\it right} =\frac{ab}{|b|^{2}},\quad \left(\frac{a}{b} \right)_\text{\it left} =\frac{ba}{|b|^{2}}, \] 
$b$ being another quaternion. Second, Euler-type formula for a quaternion exists:
$a=|a|\exp\left(\frac{y_k}{y}{\bf q}_k\cdot\theta\right)$,
a series with $y\equiv \sqrt{y_{k} y_{k} } $, $\theta \equiv \arccos \frac{x}{|a|} $. With these properties the quaternions are proved to compose the last in dimension associative division algebra over the real numbers (finite-dimensional division ring). The simplest (canonical) representation of the Q-units is given by the $2\times 2$ Pauli matrices multiplied by $-i$:
\be \label{eq2}
%\scriptsize
I=\begin{pmatrix} {1} & {0} \\ {0} & {1}\end{pmatrix},\;
{\bf q}_{1}=-i\begin{pmatrix} {0} & {1} \\ {1} & {0}\end{pmatrix},\;
{\bf q}_{2}=-i\begin{pmatrix} {0} & {-i}\\ {i} & {0}\end{pmatrix},\; {\bf q}_{3} ={\bf q}_{1} {\bf q}_{2}=
-i\begin{pmatrix} {1} & {0} \\ {0} & {-1} \end{pmatrix}
\ee
but the representation \eqref{eq2} is not unique. Another one is obtained if the imaginary (scalar) unit $i$ is replaced by matrix ${\bf q}_{2}$ of Eq.\eqref{eq2}, also an imaginary (but vector) unit, and each matrix components equal to 1 is replaced by unit $2\times 2$-matrix; then the quaternion units are described by $4\times 4$-matrices with real but still constant components. Drastic change occurs when the units acquire variable parameters. One straightforwardly shows [9,~10] that the multiplication table \eqref{eq1} keeps it form under two types of transformation of the units, each forming a group. The first is the special orthogonal group $SO(3,\mathbb{C})$ represented by $3\times 3$-matrices $O_{k'n}$ performing rotations of the vector units (at in general complex angles $\Phi $)
\be\label{eq3}
{\bf q}_{k'} = O_{k'n}(\Phi)\, {\bf q}_{n},
\ee
an irreducible representation of the group's element is an ``elementary rotation'' at angle $\Phi$ about e.g. vector ${\bf q}_{1}$
\be\label{eq4}
O_{k'n} =\left(\begin{array}{ccc} {1} & {0} & {0} \\ {0} & {\cos \Phi } & {\sin \Phi } \\ {0} & {-\sin \Phi } & {\cos \Phi } \end{array}\right).
\ee
The second group evidently leaving the table \eqref{eq1} form-invariant is the special linear group represented by matrices $U$
\be\label{eq5}
{\bf q}_{k'} = U {\bf q}_{k} U^{-1}.\ee
For the case of $2\times 2$-matrix representation of the Q-units this group is the spinor group $SL(2, \mathbb{C})$ known to be 2:1 isomorphic to $SO(3, \mathbb{C})$. An $SL(2, \mathbb{C})$ transformation operator leading to the same rotation as the operator in Eq.\eqref{eq4} is 
\be\label{eq6}
U=\left(\begin{array}{cc} {\cos \frac{\Phi}{2}} & {-i\, \sin \frac{\Phi }{2} } \\ {-i\, \sin \frac{\Phi }{2} } & {\cos \frac{\Phi }{2} } \end{array}\right) ,\;\; U^{-1} =\left(\begin{array}{cc} {\cos \frac{\Phi }{2} } & {i\, \sin \frac{\Phi }{2} } \\ {i\, \sin \frac{\Phi }{2} } & {\cos \frac{\Phi }{2} } \end{array}\right).
\ee

One readily finds that any $2\times 2$-matrix $U\in SL(2, \mathbb{C})$ is a Q-number built on the basis \eqref{eq2}, e.g. the matrices \eqref{eq6} can be written as
\be\label{eq7}
U=\cos \frac{\Phi}{2} + {\bf q}_{1} \sin \frac{\Phi }{2}, \quad U^{-1} =\cos \frac{\Phi }{2} - {\bf q}_{1} \sin \frac{\Phi }{2},
\ee 
the transformations \eqref{eq4}, \eqref{eq6} leaving vector ${\bf q}_{1}$ intact. The both groups $SO(3, \mathbb{C})$, $SL(2, \mathbb{C})$ do not change the scalar unit. It is also known that the group $SO(3, \mathbb{C})$ is 1:1 isomorphic to the Lorentz group, and that is where the relativity theory mentioned in ref. [6] appears. But the goal of this paper is the spinor structure lying in the basement of the quaternion algebra.

\section{Theory of matrices and spinor basement of Q-numbers}

In order to reveal the mathematical basement of any set of the Q-unit triads (frequently associated with three dimensions of physical space) make a useful exercise from the matrix algebra, namely, analyze the eigenvector-eigenvalue problem for an arbitrary $2\times 2$-matrix $A$ with in general complex-number elements. 

It is straightforwardly proved that \textit{A} can be represented as a sum of a diagonal matrix with non-zero trace and a traceless matrix, or as a (bi-)quaternion with scalar and vector parts (e.g. [9])
\be\label{eq8}
A=\frac{{\rm Tr}\,A}{2} \cdot I+\sqrt{\det A-\frac{{\rm Tr}^{2} A}{4}} \cdot \textbf{q};
\ee
here $I$ is the unit matrix, \textbf{q} is a traceless $2\times 2$-matrix with $\det \textbf{q}=1$, hence it is an imaginary Q-unit $\textbf{q}^{2} =-1$ belonging to some Q-triad; eigenvectors of $\textbf{q}$ surely fit for $I$. Consider two complimentary cases:

\underbar{Case (i)}: let $A$ be invertible, i.e. $\det A\neq 0$. Then Eq.\eqref{eq8} yields expression for the eigenvalues $\alpha$ of $A$
\[
\alpha =\sqrt{\det A} (\cos \Phi+\lambda \sin \Phi),\;
\cos \Phi\equiv \frac{{\rm Tr}\,A}{2\sqrt{\det A}},\]
\[{\rm \{ det}A{\rm ,\; Tr}A,{\rm \Phi \} }\in \mathbb{C}\quad\text{(the set of complex numbers)},\]
where $\lambda$ are eigenvalues of $\bf q$. According to the theory of matrices all matrices $\bf q$ are \textit{similar} [due to Eq.\eqref{eq5}], so their eigenvalues coincide with those of the canonical matrix ${\bf q}_{3} $ [from Eq.\eqref{eq2}] that obviously has $\lambda=\pm i$. Therefore 
\be\label{eq9}
\alpha^{\pm} =\sqrt{\det A} \exp (\pm i \Phi );
\ee
the same result of course follows from the solution of characteristic equation for $A$. Thus any invertible $2\times 2$-matrix $A$ has exactly two distinct eigenvalues, so it is \textit{simple},\textit{ }and due to the \textit{spectrum theorem} (e.g. [11]) can be decomposed into orthogonal parts 
\be\label{eq10}
A=\alpha ^{+} C^{+}+\alpha^{-} C^{-}, \quad C^{+} C^{-} =0, \ee
where $C^{\pm } $ are projectors formed as direct products of right (2D-column $\psi^{\pm}$) and left (2D-row $\varphi^{\pm}$) bi-orthonormal eigenvectors of \textit{A} 
\be\label{eq11}
C^{\pm } =\psi^{\pm}\varphi^{\pm}.
\ee
The projectors are known to be singular ($\det C^{\pm } =0$) idempotent ($C^{\pm } C^{\pm } =C^{\pm } $) matrices with unit trace ${\rm Tr}\, C^{\pm } =1$ equal to the only one eigenvalue of $C^{\pm }$.  

\underbar{Case (ii)}: let matrix $A$ be non-invertible (singular: $\det A\neq 0$), and let $A\rightarrow C^{\pm } $, then Eq.\eqref{eq8} yields 
\be\label{eq12}
C^{\pm }=\frac{I\pm i\textbf{q}}{2}.\ee
From Eqs. \eqref{eq11}, \eqref{eq12} one immediately obtains equivalents of the decomposition \eqref{eq10} ($A=I$ or $A=q$)
\be
I=C^{+} +C^{-} =\psi ^{+} \varphi ^{+} +\psi ^{-} \varphi ^{-}\tag{13a}
\ee
\be
q=iC^{+} -iC^{-} =i(\psi ^{+} \varphi ^{+} -\psi ^{-} \varphi ^{-} ).\tag{13b}\ee \setcounter{equation}{13}
Eqs.(13) are fundamental relations demonstrating that units of any Q-triad consist of more elementary objects constituting a bi-orthonormal basis $\{\varphi^{\pm},~\psi ^{\pm}\}$ of a 2D-vector space, (in general) a complex-number-valued surface. On the one hand, comparison of Eqs.(13) and \eqref{eq5} shows that the vectors of the basis are spinors whose transformations $\psi'=U\psi $, $\varphi'=\varphi \, U^{-1} $, $U\in SL(2, \mathbb{C})$ do not affect quaternion multiplication rule \eqref{eq1}. On the other hand, the vectors$\, \, \psi ^{\pm } $ and co-vectors $\varphi ^{\pm } $ obey the standard ``metric requirements''
$\varphi^{\pm }\psi ^{\pm } =1$, $\varphi ^{\pm }\psi ^{\mp } =0$. In dyad notations (as Lam\'e coefficients linking basic and tangent 2D-surfaces) 
\begin{equation}
\label{eq14}
\psi^{\pm }\equiv h_{(M)}^{A},\quad
\varphi ^{\pm } \equiv h_{(N)B}
\end{equation} 
were $A, B, ...=1,2$ are contravariant and covariant component indices, and $(M),(N),...=1,2$ are the parity indices (always written at the bottom of a symbol), the orthogonality and normalization conditions shrink to the unique equation
\begin{equation} \label{eq15}
h_{(M)}^{A} \, h_{(N)A} =\delta _{MN}
\end{equation} 
while Eq.(13a) acquires the form
\be\label{eq16}
h_{(M)}^{A} h_{(M)B} =\delta _{B}^{A},\ee
$\delta _B^A$, $\delta_{MN}$ being the 2D Kronecker symbols for different types of indices. Eq.\eqref{eq15} describes the metric of 2D-tangent plane, while expressions for metric tensor of the 2D-base and its reciprocal follow from Eq.\eqref{eq16}
\be\label{eq17}
g_{AB}\equiv h_{(M)A} h_{(M)B},\quad
g^{BC}\equiv h_{(M)}^{B} h_{(M)}^{C}.\ee

Now return to the simplest case of Q-units given by Eq.\eqref{eq2} and notice that the constant (``plane'') spinors of the form 
\begin{equation} \label{eq18}
\psi^{+} =\left(\begin{array}{c} 0 \\ 1 \end{array}\right),\;
\varphi^{+} =\left(0,\; 1\right),\;
\psi^{-} =\left(\begin{array}{c} 1 \\ 0 \end{array}\right),\;
\varphi^{-} =\left( 1,\;  0 \right)
\end{equation} 
are the eigenvector of ${\bf q}_{3} $ [of Eq.\eqref{eq2}], hence they constitute this unit according to Eq.(13b)
\be\label{eq19}
{\bf q}_3 =i (\psi^{+} \varphi^{+}-\psi^{-}\varphi^{-} ),\ee
Eq.(13a) holding as well. But, one easily finds that the representation \eqref{eq18} enables one to build all other vector Q-units of the set \eqref{eq2} composing linear combination of tensor products of the spinors with opposite parity 
\be\label{eq20}
{\bf q}_2 =\psi^{+} \varphi^{-} -\psi^{-} \varphi^{+},
\ee
\be\label{eq21}
{\bf q}_1 =-i(\psi^{+} \varphi^{-} +\psi^{-} \varphi^{+} ).
\ee

An important observation must be made here. The Q-units given by Eqs.(13a), \eqref{eq19}--\eqref{eq21} obey the basic multiplication rule \eqref{eq1} \textit{irrespectively} of specific form of the spinor elements entering their structure. One directly verifies the fact composing all possible products of the units. Moreover, it is shown in the paper [12] that the representation of the Q-units through quadratic combinations of spinors is unique since only two linear combinations of nilpotent matrices of the type $\psi ^{\pm } \varphi ^{\mp } $ and only two linear combinations of idempotent matrices of the type $\psi ^{\pm } \varphi ^{\pm } $ exist; namely these four combinations form four Q-units. This means that any pair of orthogonal normalized spinor functions, not only those given by Eqs.\eqref{eq18}, form a set of Q-units, and mathematical properties of the spinors fully determine geometric behavior of the respective Q-triad (in particular its behavior as a relativistic frame of reference what may have useful physical consequences).

\section{Eigenvectors of different Q-units and cyclic recurrent formulae}

It is shown above that each Q-unit has its spinor eigenvectors (with the same eigenvalues $\pm i$), and any set of Q-units can be built using the only one set of spinor-vectors as e.g. in Eqs. (13a), \eqref{eq19}--\eqref{eq21}. This fact means that all eigenvectors of any Q-triad are functionally dependent. A special example of respective relations is adduced in [12]; below the relations are obtained in the universal procedure and in the general form. 

Let the spinors $\rho ^{\pm }$, $\xi ^{\pm }$ be left and right eigenvectors of ${\bf q}_{1}$ and $\eta ^{\pm}$, $\theta ^{\pm }$ be left and right eigenvectors of ${\bf q}_{2} $ similarly to $\varphi ^{\pm } ,\, \, \psi ^{\pm } $, eigenvector of ${\bf q}_{3} $. Then the scalar unit and a ring of vector Q-units have three equivalent representations 
\be I=\xi ^{+} \rho ^{+} +\xi ^{-} \rho ^{-} =\theta ^{+} \eta ^{+} +\theta ^{-} \eta ^{-} =\psi ^{+} \varphi ^{+} +\psi ^{-} \varphi ^{-} \tag{22a}\ee
\be{\bf q}_{1} =i\, (\xi ^{+} \rho ^{+} -\xi ^{-} \rho ^{-} )\, =\theta ^{+} \eta ^{-} -\theta ^{-} \eta ^{+} =-i\, (\psi ^{+} \varphi ^{-} +\psi ^{-} \varphi ^{+} )\tag{22b}\ee
\be{\bf q}_{2} =-i\, (\xi ^{+} \rho ^{-} +\xi ^{-} \rho ^{+} )=i\, (\theta ^{+} \eta ^{+} -\theta ^{-} \eta ^{-} )=\psi ^{+} \varphi ^{-} -\psi ^{-} \varphi ^{+} \tag{22c}\ee
\be{\bf q}_{3} $$\, =$$\xi ^{+} \rho ^{-} -\xi ^{-} \rho ^{+} $$\, =$$-i\, (\theta ^{+} \eta ^{-} +\theta ^{-} \eta ^{+} )$$\, =$$i\, (\psi ^{+} \varphi ^{+} -\psi ^{-} \varphi ^{-} ) \tag{22d}.\ee
\setcounter{equation}{22}
Altogether here are 12 nonlinear equations for 12 functions
$\rho^{\pm }$, $\xi^{\pm }$; $\eta^{\pm }$, $\theta^{\pm }$; $\varphi^{\pm }$, $\psi^{\pm }$. But according to the rule \eqref{eq1} the scalar unit (22a) equals minus square of any vector unit from the set (22), while each vector unit is a product of two others, e.g. Eq.(22d) is product of Eqs.(22b) and (22c). Moreover, the three spinor couples are normalized in 6 equations and also they obey 6 orthogonality conditions, so a great degree of ``algebraic symmetry'' is expected in the relations linking the eigenvectors.

Avoiding detailed deduction of all correlations demonstrate the routine for only one spinor. For instance spinor $\xi^{+}$ is found as a function of $\theta^{\pm }$ from the first equalities of the set (22) after multiplication from the right of: Eq.(22a) by $\xi^{+}$, Eq.(22b) by ($-i\xi^{+} $), Eq.(22c) by $i\xi^{-}$, and Eq.(22d) by $\xi^{-}$; this yields the four equations
\be\xi ^{+} =(\eta ^{+} \xi ^{+} )\, \theta ^{+} +(\eta ^{-} \xi ^{+} )\, \theta ^{-}\tag{23a}\ee
\be\xi ^{+} =-i(\eta ^{-} \xi ^{+} )\, \theta ^{+} +i(\eta ^{+} \xi ^{+} )\, \theta ^{-} \tag{23b}\ee
\be\xi ^{+} =-(\eta ^{+} \xi ^{-} )\, \theta ^{+} +(\eta ^{-} \xi ^{-} )\, \theta ^{-}\tag{23c}\ee
\be\xi ^{+} =-i(\eta ^{-} \xi ^{-} )\, \theta ^{+} -i(\eta ^{+} \xi ^{-} )\, \theta ^{-}\tag{23d}.\ee
\setcounter{equation}{23}
All coefficients at the basic spinor-vector $\theta^{+}$ must be equal, the coefficients at $\theta^{-}$ must be equal too; one readily finds that the both requirements result in the only one line of equalities 
\begin{equation} \label{eq24}
\eta ^{+} \xi ^{+} =-i\eta ^{-} \xi ^{+} =
-\eta ^{+} \xi ^{-} =-i\eta ^{-} \xi ^{-}
\end{equation} 
relating all possible contractions (scalar products) of co-vectors
$\eta ^{\pm } $ from the spinor surface No.2
$\{\eta^{\pm},\,\theta^{\pm}\}$ with vector $\xi^{\pm}$ from the spinor surface No.1 $\{\rho^{\pm },\,\xi^{\pm}\}$, each scalar product in general being a complex number. Eqs.(23), \eqref{eq24} allow expressing $\xi^{+}$ through $\theta^{\pm }$ up to a constant factor, chosen here as $(\eta^{+} \xi^{+} )$
\be\label{eq25}
\xi^{+}=(\eta^{+}\xi^{+})\, (\theta^{+}+i\theta^{-}).\ee
Co-vector $\rho ^{+}$ for this spinor is found as a function of co-vectors $\eta^{\pm}$ in similar procedure after multiplication of Eqs.(22) form the left by $\pm\rho^{\pm}$ with or without factor $i$. Analogous [but ``inverse'' to those of Eq.\eqref{eq24}] correlations between all scalar products of co-vectors
$\rho^{\pm }$ from the surface No.1 with vector $\theta^{\pm }$ from the surface No.2 hold
\be\label{eq26}
\rho ^{+} \theta ^{+} =-\rho ^{-} \theta ^{+} =i\rho ^{+} \theta ^{-} =i\rho ^{-} \theta ^{-}.\ee
The sought for function is found as
\be\label{eq27}
\rho ^{+} =(\rho ^{+} \theta ^{+} )\, (\eta ^{+} +i\eta ^{-} );\ee
the constant factors of Eqs.\eqref{eq25}, \eqref{eq27} are interconnected by the normalization constraint 
\be\label{eq28}
\rho ^{+} \xi ^{+} =2(\eta ^{+} \xi ^{+} )(\rho ^{+} \theta ^{+} )=1.\ee
Similarly all other spinor relations are obtained; the full list is given below in the form: a spinor vector of the cycle $\xi^{\pm } \rightarrow \theta^{\pm } \rightarrow \psi ^{\pm }\rightarrow \xi ^{\pm }$ (co-vector of the cycle $\rho ^{\pm }\rightarrow \eta ^{\pm } \rightarrow \varphi ^{\pm } \rightarrow \rho ^{\pm } $) is expressed through ``previous'' spinor (left equality) and through ``following'' vector (right equality) 
\be(\varphi ^{+} \xi ^{+} )(\psi ^{+} -\psi ^{-} )=\xi ^{+} =(\eta ^{+} \xi ^{+} )\, (\theta ^{+} +i\theta ^{-} ),  \tag{29a}\ee
\be(\rho ^{+} \theta ^{+} )(\xi ^{+} -\xi ^{-} )=\theta ^{+} =(\varphi ^{+} \theta ^{+} )\, (\psi ^{+} +i\psi ^{-} ),  \tag{29b}\ee
\be(\rho ^{+} \xi ^{+} )(\theta ^{+} -\theta ^{-} )=\psi ^{+} =(\rho ^{+} \psi ^{+} )\, (\xi ^{+} +i\xi ^{-} ),  \tag{29c}\ee
\be-i(\varphi ^{+} \xi ^{+} )(\psi ^{+} +\psi ^{-} )=\xi ^{-} =(\eta ^{+} \xi ^{+} )\, (-\theta ^{+} +i\theta ^{-} ), \tag{29d}\ee
\be-i(\rho ^{+} \theta ^{+} )(\xi ^{+} +\xi ^{-} )=\theta ^{-} =(\varphi ^{+} \theta ^{+} )\, (-\psi ^{+} +i\psi ^{-} ),  \tag{29e}\ee
\be-i(\rho ^{+} \xi ^{+} )(\theta ^{+} +\theta ^{-} )=\psi ^{-} =(\rho ^{+} \psi ^{+} )\, (-\xi ^{+} +i\xi ^{-} );  \tag{29f}\ee
\be(\rho ^{+} \psi ^{+} )(\varphi ^{+} -\varphi ^{-} )=\rho ^{+} =(\rho ^{+} \theta ^{+} )\, (\eta ^{+} -i\eta ^{-} ),  \tag{29g}\ee
\be(\eta ^{+} \xi ^{+} )(\rho ^{+} -\rho ^{-} )=\eta ^{+} =(\eta ^{+} \psi ^{+} )\, (\varphi ^{+} -i\varphi ^{-} ),  \tag{29h}\ee
\be(\varphi ^{+} \theta ^{+} )(\eta ^{+} -\eta ^{-} )=\varphi ^{+} =(\varphi ^{+} \xi ^{+} )\, (\rho ^{+} -i\rho ^{-} ),  \tag{29i}\ee
\be i(\rho ^{+} \psi ^{+} )(\varphi ^{+} +\varphi ^{-} )=\rho ^{-} =(\rho ^{+} \theta ^{+} )\, (-\eta ^{+} -i\eta ^{-} ),  \tag{29j}\ee
\be i(\eta ^{+} \xi ^{+} )(\rho ^{+} +\rho ^{-} )=\eta ^{-} =(\eta ^{+} \psi ^{+} )\, (-\varphi ^{+} -i\varphi ^{-} ),  \tag{29k})\ee
\be i(\varphi ^{+} \theta ^{+} )(\eta ^{+} +\eta ^{-} )=\varphi ^{-} =(\varphi ^{+} \xi ^{+} )\, (-\rho ^{+} -i\rho ^{-} );  \tag{29l}\ee
constant factors (in common form of contraction ``positive co-vector with positive vector'') are distinguished from Eqs.\eqref{eq24}, \eqref{eq26} and the remaining four equality lines
\be\varphi ^{+} \theta ^{+} =-i\varphi ^{-} \theta ^{+} =-\varphi ^{+} \theta ^{-} =-i\varphi ^{-} \theta ^{-}, \tag{30a}\ee
\be \eta ^{+} \psi ^{+} =-\eta ^{-} \psi ^{+} =i\eta ^{+} \psi ^{-} =i\eta ^{-} \psi ^{-},\tag{30b}\ee
\be\rho ^{+} \psi ^{+} =-i\rho ^{-} \psi ^{+} =-\rho ^{+} \psi ^{-} =-i\rho ^{-} \psi ^{-},\tag{30c}\ee
\be\varphi ^{+} \xi ^{+} =-\varphi ^{-} \xi ^{+} =i\varphi ^{+} \xi ^{-} =i\varphi ^{-} \xi ^{-}.\tag{30d}\ee
It is important to note that the normalization condition Eq.\eqref{eq28} fit as well for coupling spinors of ``negative'' parity $\rho ^{-} \xi ^{-} =1$, and two additional normalization constraints should hold 
\be\eta ^{\pm } \theta ^{\pm } =2(\eta ^{+} \psi ^{+} )(\varphi ^{+} \theta ^{+} )=1,\tag{31a}\ee
\be\varphi ^{\pm } \psi ^{\pm } =2(\varphi ^{+} \xi ^{+} )(\rho ^{+} \psi ^{+} )=1,\tag{31b}\ee
while all orthogonality conditions of the type $\rho ^{\pm } \xi ^{\mp }=0$ \textit{are satisfied identically}. Due to Eqs. \eqref{eq26}, \eqref{eq28}, (30), (31) only 3 scalar products can be now considered independent; chose the three contractions and denote  them as 
\setcounter{equation}{31}
\be\label{eq32}
\rho ^{+} \theta ^{+} \equiv x,\quad \eta ^{+} \psi ^{+} \equiv y, \quad \varphi ^{+} \xi ^{+} \equiv z.\ee

As expected the obtained equalities demonstrate a noticeable algebraic symmetry. Indeed, Eqs.(29a,b,c) expressing positive spinor-vectors through a previous one and a following one have similar forms; the same property possess other three groups of spinors. One also notes that the subsequent substitution of spinors
$\xi^{\pm}(\psi^{\pm})\rightarrow\theta^{\pm}[\xi^{\pm}(\psi^{\pm})]\rightarrow\psi^{\pm}\{\theta^{\pm}[\xi^{\pm}(\psi^{\pm})]\}$ within Eqs.(29a--f) yields definite value of the cubic product 
\be\label{eq33}
(\rho ^{+} \theta ^{+} )(\eta ^{+} \psi ^{+} )(\varphi ^{+} \xi ^{+} )\equiv xyz=\frac{1-i}{4};
\ee
it can be easily verified that the rest of Eqs.(29) gives the same result. So only two contractions out of three from Eq.\eqref{eq32} are in fact independent, hence two free coefficients at two different spinor-vectors (or co-vectors) exhaustively determine all other spinors. The above observations hint to construct a recurrent formula for a ``following'' spinor from a ``previous'' one; it is convenient to use for the purpose the notation \eqref{eq14} of spinors as dyads but endowed with an extra index referring a spinor to one of the three spinor sets 
\be\label{eq34}
{}^{n}\! h_{(M)}^{A} \equiv (\xi ^{\pm } ,\theta ^{\pm } ,\psi ^{\pm } ),\quad {}^{n}\! h_{(N)B} \equiv (\rho ^{\pm } ,\eta ^{\pm } ,\varphi ^{\pm } ),\quad n=1,\, 2,\, 3,\ee
so that ${}^{1}\! h_{(M)}^{A} \equiv \xi ^{\pm}$, $^{2}\! h_{(M)}^{A} \equiv \theta ^{\pm}$, $^{1}\! h_{(N)B} \equiv \rho ^{\pm }$, etc; then the left equalities of Eq.(29b, e) are written as
\[{}^{2}\! h_{(1)}^{A} =({}^{1}\! h_{(1)B} \,{}^{2}\! h_{(1)}^{B} \, )({}^{1}\! h_{(1)}^{A} -{}^{1}\! h_{(2)}^{A} ), \] 
\[{}^{2}\! h_{(2)}^{A} =-i\, ({}^{1}\! h_{(1)B}\, {}^{2}\! h_{(1)}^{B} \, )({}^{1}\! h_{(1)}^{A} +{}^{1}\! h_{(2)}^{A} ).\] 
These separate equations can be united into the following one (for $n=1$)
\begin{equation} \label{eq35}
{}^{n+1}\! h_{(K)}^{A} =\frac{1+i^{2K-1} }{2} \left({}^{n}\! h_{(M)B}\, {}^{n+1}\! h_{(M)}^{B} \right)\, (\delta _{NK} +\varepsilon _{NK} )\; {}^{n}\! h_{(N)}^{A},
\end{equation} 
where $\varepsilon_{NK} \equiv \delta _{N}^{1} \delta _{K}^{2} -\delta _{N}^{2} \delta _{K}^{1} $ is 2D symbol of Levi-Civita, summation is implied in all repeating 2D indices, and relation of the type
\begin{equation} \label{eq36}
\rho ^{+} \theta ^{+} =\frac{1+i}{2} (\rho ^{+} \theta ^{+} +\rho ^{-} \theta ^{-} )\equiv \frac{1+i}{2} \, \left({}^{1}\! h_{(M)B}\, {}^{2}\! h_{(M)}^{B} \right)\equiv x
\end{equation} 
is used showing implied structure of a freely chosen coefficient $x$. The recurrent formula \eqref{eq35} defines functional dependence between any spinor vectors; its ``cyclic'' use\footnote{ If  $n=3$ , then  $n+1\rightarrow 1$.} leads to Eqs.(29a--f). Recurrent relations for spinor co-vectors, corresponding to the rest of the set (29g--l), are obtained from Eq.\eqref{eq35} by lowering vector index with the help of respective 2D metric tensor. Such a metric for, e.g., the spinor surface No.2 $\{\eta^{\pm},\, \theta ^{\pm}\}$ is defined as in Eq.\eqref{eq17}
\be\label{eq37}
{}^{2}\! g(\eta)\equiv \eta ^{+} \eta ^{+} +\eta ^{-} \eta ^{-}\quad
\Longleftrightarrow \quad ^{2}\! g_{AC} ={}^{2}\! h_{(M)A}\,{}^{2}\! h_{(M)C}.\ee

But if this metric is regarded from the viewpoint of spin surface No.1 $\{\rho ^{\pm }, \xi ^{\pm}\}$ the spinors $\eta ^{\pm } $ should be substituted by $\rho ^{\pm } $ with the help of Eqs.(29h,k), then Eq.\eqref{eq37} reads as
\[
{}^{2}\! g(\rho )\equiv -2(\eta ^{+} \xi ^{+} )^2
(\rho ^{+} \rho ^{-} +\rho ^{-} \rho ^{+} ).
\] 
In the dyad notations [using Eq.\eqref{eq28}, \eqref{eq36}] the metric has the form
\[
{}^{2} g_{AC} =
\frac{i}{\left({}^{1} h_{(L)D}\;\, {}^{2} h_{(L)}^{D} \right)^2}
\;\sigma_{PQ}\, {}^{1}\! h_{(P)A} {}^{1}\!h_{(Q)C},\;\;\quad
\sigma_{PQ} \equiv \delta _{P}^{1} \delta _{Q}^{2} +\delta _{P}^{2} \delta _{Q}^{1},
\] 
what helps to suggest its ``recurrent format'' for any spinor surface
\be\label{eq38}
{}^{n+1}\!g_{AC} ({}^{n}\!h)=\frac{i}{\left({}^{n}\! h_{(L)D}\;\, {}^{n+1}\!h_{(L)}^{D} \right)^{2} } \; \sigma_{PQ}\, {}^{n}\!h_{(P)A}\, {}^{n}\! h_{(Q)C}\ee

The metric \eqref{eq38} lowers upper index of the spinor-vector \eqref{eq35} yielding the cyclic recurrent formula for spinor-co-vector
\be\label{eq39}
{}^{n+1}\!h_{AC} =\frac{i(1+i^{2K-1} )}{2\left({}^{n}\! h_{(L)D}\; {}^{n+1}\! h_{(L)}^{D} \right)} \, (\sigma _{PQ} -\tau _{{\rm PQ}} )\,{}^{n}\!h_{(Q)C}, \quad\;  \tau _{PQ} \equiv \delta _{P}^{1} \delta _{Q}^{1} -\delta _{P}^{2} \delta _{Q}^{2};\ee

Eqs. \eqref{eq39} lead to correlations (29g--l). Thus all spinor-vector (co-vector) links \eqref{eq35}, \eqref{eq39} are established and generic expressions for metric tensors of any spinor surface \eqref{eq38} as functions of any spinor families are deduced. 

\section{Table of spinor scalar products}

Basic spinors' orthonormality conditions are ``scalar products'' on the dyad's proper spin-surface. The correlations between all other scalar products are given by Eqs.\eqref{eq24}, \eqref{eq26}, (30) together with imposed conditions \eqref{eq28}, (31). Taking into account this information and using notations \eqref{eq32} form a square table of all possible scalar products ``co-vector $\times$ vector'' placed at the intersection of rows (co-vectors) and columns (vectors)\\[0.5em]

\underbar{Table 1}.\nopagebreak[4]\\[0.2em]

%\begin{tabular}{|p{0.3in}|p{0.4in}|p{0.4in}|p{0.4in}|p{0.4in}|p{0.4in}|p{0.4in}|} \hline
\nopagebreak[4]
\begin{tabular}{|c|c|c|c|c|c|c|} \hline
& $\xi ^{+}$ & $\theta ^{+} $ & $\psi ^{+} $ & $\xi ^{-} $ & $\theta ^{-} $ & $\psi ^{-} $ \\ \hline 
$\rho ^{+} $ & $1$ & $x$ & $  \displaystyle\frac{1}{2z} $ & 0 & $-ix$ & $\displaystyle-\frac{1}{2z} $ \\ \hline 
$\eta ^{+} $ & $\displaystyle\frac{1}{2x} $ & $1$ & $y$ & $\displaystyle-\frac{1}{2x} $ & 0 & $-iy$ \\ \hline 
$\varphi ^{+} $ & $z$ & $\frac{1}{2y} $ & $1$ & $-iz$ & $\displaystyle-\frac{1}{2y} $ & 0 \\ \hline 
$\rho ^{-} $ & 0 & $-x$ & $\displaystyle\frac{i}{2z} $ & $1$ & $-ix$ & $\displaystyle\frac{i}{2z} $ \\ \hline 
$\eta ^{-} $ & $\displaystyle\frac{i}{2x} $ & 0 & $-y$ & $\displaystyle\frac{i}{2x} $ & $1$ & $-iy$ \\ \hline 
$\varphi ^{-} $ & $-z$ & $\displaystyle\frac{i}{2y} $ & 0 & $-iz$ & $\displaystyle\frac{i}{2y} $ & $1$ \\ \hline 
\end{tabular}

\vskip1em

The Table 1 is obviously composed by four blocs of square $3\times 3$-matrices of the type
\be\label{eq40}
T\equiv \left(\begin{array}{cc} {S^{+} } & {S^{\pm} } \\ {S^{\mp} } & {S^{-}} \end{array}\right),\ee
where elements of the blocs $S^{+},\ S^{{-}}$ are scalar products of spinors of the same parity, elements of $S^{\pm},\ S^{\mp}$ are scalar products of spinors of opposite parity. Analyzing properties of the matrix \eqref{eq40} one discovers that determinants of all its blocs of vanish
\be\label{eq41}
\det S^{+}=\det S^{-}=\det S^{\pm}=\det S^{\mp}=0,\ee 
though not identically, but due to interdependence of the free factors \eqref{eq33}; so Eq.\eqref{eq33} represents a special solution of Eqs.\eqref{eq41}. The Table 1 represents a set of fundamental relations between quaternion spinors in general case, but in applications particular values of the scalar products can be needed. Table 2 placed below reflects a special case where explicitly showed spinor matrices are eigenvectors of the simplest vector Q-units from Eqs.\eqref{eq2}, while coefficients satisfying the condition \eqref{eq33} are $\displaystyle x=\frac{1-i}{2}$, $y=z=\displaystyle\frac{1}{\sqrt{2} }$.  

\underbar{Table 2}.\\[0.2em]

%\begin{tabular}{|p{0.8in}|p{0.5in}|p{0.5in}|p{0.5in}|p{0.5in}|p{0.5in}|p{0.5in}|} \hline 
\begin{tabular}{|l|c|c|c|c|c|c|} \hline
& $\!\!\scriptstyle\xi ^{+} =\frac{1}{\sqrt{2} } \begin{pmatrix} {-1} \\ {1} \end{pmatrix}\!\!\!\!\!\!$
& $\!\!\scriptstyle\theta ^{+} =\frac{1}{\sqrt{2} } \begin{pmatrix} {i} \\ {1} \end{pmatrix}\!\!\!\!\!\!$
& $\!\!\scriptstyle \psi ^{+} =\begin{pmatrix} {0} \\ {1} \end{pmatrix}\!\!\!\!\!\!$
& $\!\!\scriptstyle \xi ^{-} =\frac{-i}{\sqrt{2} } \begin{pmatrix} {1} \\ {1} \end{pmatrix}\!\!\!\!\!\!$
& $\!\!\scriptstyle \theta ^{-} =\frac{1}{\sqrt{2} } \begin{pmatrix} {i} \\ {-1} \end{pmatrix}\!\!\!\!\!\!$
& $\!\!\scriptstyle \psi ^{-} =\begin{pmatrix} {1} \\ {0} \end{pmatrix}\!\!\!\!\!\!$ \\ \hline 
$\!\!\rho ^{+} =\frac{1}{\sqrt{2} }(-1,\ 1)\!\!\!\!$ & $1$ & $\displaystyle\frac{1-i}{2} $ & $\displaystyle\frac{1}{\sqrt{2} } $ & $0$ & $\displaystyle-\frac{1+i}{2} $ & $\displaystyle-\frac{1}{\sqrt{2} } $ \\ \hline 
$\!\!\eta^{+} =\frac{1}{\sqrt{2} } ( -i,\ 1)\!\!\!\!$ & $\displaystyle\frac{1+i}{2} $ & $1$ & $\displaystyle\frac{1}{\sqrt{2} } $ & $\displaystyle-\frac{1+i}{2} $ & $0$ & $\displaystyle-\frac{i}{\sqrt{2} } $ \\ \hline 
$\!\!\varphi^{+} =(0,\ 1)\!\!\!\!\!\!$ & $\displaystyle\frac{1}{\sqrt{2} } $ & $\displaystyle\frac{1}{\sqrt{2} } $ & $1$ & $\displaystyle-\frac{i}{\sqrt{2} } $ & $\displaystyle-\frac{1}{\sqrt{2} } $ & $0$ \\ \hline 
$\!\!\rho ^{-}=\frac{i}{\sqrt{2} } (1,\ 1)\!\!\!\!$ & $0$ & $\displaystyle-\frac{1-i}{2} $ & $\displaystyle\frac{i}{\sqrt{2} } $ & $1$ & $\displaystyle-\frac{1+i}{2} $ & $\displaystyle\frac{i}{\sqrt{2} } $ \\ \hline 
$\!\!\eta^{-} =-\frac{1}{\sqrt{2} } (i,\ 1)\!\!\!\!$ & $\displaystyle-\frac{1-i}{2} $ & $0$ & $\displaystyle-\frac{1}{\sqrt{2} } $ & $\displaystyle-\frac{1-i}{2} $ & $1$ & $-iy$ \\ \hline 
$\!\!\varphi ^{-} =(1,\ 0)\!\!\!\!\!$ & $-\frac{1}{\sqrt{2} } $ & $\displaystyle\frac{i}{\sqrt{2} } $ & $0$ & $\displaystyle-\frac{i}{\sqrt{2} } $ & $\displaystyle\frac{i}{\sqrt{2} } $ & $1$ \\ \hline 
\end{tabular}

\vskip1em
If the scalar products in the Table 2 are traditionally treated as mutual projections of involved unit vectors (i.e. as cosines of an angle between the vectors) then one has to conclude that all vectors from different spin-surfaces are inclined to each other at an angle proportional (up to a degree of the imaginary unit) to $\pi /4$; it is readily verified that any other choice of the Q-units' eigenvectors and the coefficients \eqref{eq32} will lead to the similar result. 

\section{Discussion}

Thus from the most general positions a ``structural content'' of any sets of quaternion units is revealed, and fundamental properties of the constituent functions are established. It is shown that the functions have a dual mathematical meaning: they are spinors (i.e. objects of 1/2 dimension of a vector) and in the same time, born in couples, they form dyads, basic vectors of 2D-surfaces (called here spin-surfaces). There are other characteristic properties of the spinors to be emphasized. The vector Q-units are interrelated by a non-linear action, vector multiplication, while respective spinor functions (``square roots'' of the units') are linearly dependent, and a couple of spinor (2 rows and 2 columns of `$\pm $' parity), eigenvectors of any Q-triad's vector, is sufficient for building the whole set of Q-units. Moreover, each set of Q-units is expressed through quadratic combinations of one spinor couple uniquely. Hence altogether there are three variants of expressing all Q-units through spinor couples born as eigenvectors by three different vector units. With the help of these expressions algebraic interrelations between spinor sets from the same family are found, and a table of all possible scalar products of the spinors is produced. 

The question arises: what for is this boring math routine? To understanding of the author, the results of the study promise to reveal in the spinor algebra ``non less geometry'' than in a Q-triad naturally related to a Cartesian coordinate system in a 3D space (locally quasi-Euclidean\footnote{This local (tangent) 3D space has Euclidian plane metric tensor  $g_{kn} =\delta _{kn} $ , but it is not ``completely'' Euclidian since vector multiplication in it is not commutative.}). The considered above Q-spinors, regarded as vectors and co-vectors on spin-surfaces evidently form a certain mathematical entity that can be named ``pre-geometry''\footnote{ In a sense this model of pre-geometry corresponds to the famous notion suggested by J. A. Wheeler.}, if the term ``geometry'' is reserved for properties of the physical space admitting experimental observation and measurement (by vector Q-units in this context). But if geometrical perception of a Q-triad meets no obstacles, the visualization of spinor pre-geometry constituting the basement of 3D space dimensions still remains vague. An advance in this domain seems a challenging task because construction of a plausible geometric image of ``hardly-conceivable'' spinor could lead to a better understanding of the geometro-physical foundations of quantum mechanics. Indeed, Eqs.\eqref{eq9}, \eqref{eq10} evoke existence of the spinor function 
\be\label{eq42}
\Psi \equiv \sqrt{\det A}\, \exp (i\Phi)\psi.
\ee
that comprises all ingredients of de Broglie-Pauli-type quantum mechanical wave function: it contains a complex-number valued amplitude ${\rm det}A\in \mathbb{C}$, a pure wave part $\exp (\pm i\Phi)$ (if $\Phi \in \mathbb{R}$ what can be always achieved), and a spin-term $\psi ^{+} $ or $\psi ^{-} $; each constituent may depend on space-time coordinates. In the same time Eq.\eqref{eq42} describes components of a 2D vector belonging to a pre-geometric spin-surface locally determined by the basic vector couple $h_{(M)}^{A} \equiv \psi ^{\pm } $. It is worthwhile to add that a picture of an abstract spin-surface given separately of its native Q-triad would be hardly informative, hence hardly helpful. The pre-geometry obviously ``interact'' with the geometry, e.g. rotations of $SU(2)$-spinor should cause double-angled rotation of its Q-triad, and this even must be reflected by the mutual image. The Q-spinor properties detailed in this study seem to be a good basement for an attempt to generate such an image; a variant of consistent picture will be offered in a following publication.


\begin{thebibliography}{99}
\bibitem{1} A.P.Yefremov, Adv.Sci.Lett. (USA), V.3, P. 537--542 (2010).

\bibitem{2} W. R. Hamilton, Lectures on Quaternions, Hodges and Smith, Dublin (1853).

\bibitem{3} R. Fueter, Comm. Math. Helv. V4, P. 9--20 (1932).

\bibitem{4} A.P. Yefremov,  Lett. Nuovo Cim. V37(8), P.315--316 (1983).

\bibitem{5} A.P.Yefremov, F. Smarandache, V. Christianto, Progress in Physics, V3, P. 42--50 (2007).

\bibitem{6} A.P.Yefremov, ``Six-Dimensional Rotational Relativity'', Acta Phys. Hun. V11(1-2), P. 147--153 (2000).

\bibitem{7} A.P. Yefremov, Adv.Sci.Lett.V1, P. 179--186 (2008).

\bibitem{8} P.Rastall, Rev. Mod. Phys., V2, P. 820--832 (1964).

\bibitem{9} A.P.Yefremov, Hypercomplex Numbers in Geometry and Physics, V1, P.104--119 (2004).

\bibitem{10} A.P.Yefremov, Quaternion Spaces, Frames and Physical Fields, Moscow, RUDN Publ. (2005), P.41.

\bibitem{11} P.Lancaster, M.Tismenetsky, The Theory of Matrices, Second Edition with Applications, Academic Press, San Diego, USA, London, UK (1985), P. 154.

\bibitem{12} A.P. Yefremov, Adv.Sci.Lett. V5, P. 288--293 (2012).

%\bibitem{13} A.P.Yefremov, ``Structure of Hypercomplex Units and Exotic Numbers as Sections of B-Quaternions'', Adv.Sci.Lett. (USA), V.3, P. 537--542 (2010)

%\bibitem{14} W. R. Hamilton, Lectures on Quaternions, Hodges and Smith, Dublin (1853).

%\bibitem{15} R. Fueter, ``Analytische Funktionen einer Quaternionenvariablen``, Comm. Math. Helv. V4, P. 9--20 (1932).

%\bibitem{16} A.P. Yefremov,  ``Quaternionic multiplication rule and a local Q-metric'', Lett. Nuovo Cim. V37\textbf{ }(8), P. 315--316 (1983).

%\bibitem{17} A.P.Yefremov, F. Smarandache, V. Christianto, ``Yang-Mills field from quaternion space geometry, and its Klein-Gordon representation'', Progress in Physics, V3, P. 42--50 (2007).

%\bibitem{18} A.P. Yefremov, ``Quaternion Model of Relativity: Solutions for Non-Inertial Motions and New Effects'', Adv.Sci.Lett. V1, P. 179--186 (2008)

%\bibitem{19} A.P.Yefremov, ``Six-Dimensional Rotational Relativity'', Acta Phys. Hun. V11(1-2), P. 147--153 (2000).

%\bibitem{20} P.Rastall, ``Quaternions in Relativity'', Rev. Mod. Phys., V2, P. 820--832 (1964)

%\bibitem{21} A.P.Yefremov, ``Quaternions: Algebra, Geometry and Physical Theories'', Hypercomplex Numbers in Geometry and Physics, V1, P.104-119 (2004).

%\bibitem{22} A.P.Yefremov, ``Quaternion Spaces, Frames and Physical Fields'', Moscow, RUDN Publ. (2005), P. 41.

%\bibitem{23} P.Lancaster, M.Tismenetsky, The Theory of Matrices, Secons Edition with Applications, Academic Press, San Diego, USA, London, UK (1985), P. 154

%\bibitem{24} A.P. Yefremov, ``Splitting of 3D Quaternion Dimensions into 2D Sells and a 'Worlds Screen Surface Geometry' '', Adv.Sci.Lett. V5, P. 288--293 (2012)

\end{thebibliography}
\end{document}